\documentclass[twocolumn,pra,showpacs, nofootinbib]{revtex4}

\usepackage{amssymb}
\usepackage{amsmath}
\usepackage{hyperref}
\usepackage{graphicx}
\usepackage{wrapfig}
\usepackage{youngtab}

\newcommand{\ket}[1]{\left|#1\right\rangle}

\newcommand{\braket}[2]{\left\langle#1\right|\left.#2\right\rangle}

\begin{document}

\title{Comparison between continuous wave and pulsed laser EQKD Systems}

\author{Patrick Rice}
\email{prrice@lanl.gov}

\author{Israel J. Owens}
\email{iowens@lanl.gov}

\affiliation{Applied Modern Physics (P-21), MS D454, Los Alamos National Laboratory, Los Alamos, NM 87545, USA}

\begin{abstract}

Advances in quantum physics and computational complexity threaten the security of present day cryptographic systems and have driven the development of quantum key distribution (QKD). Entangled quantum key distribution (EQKD) is a secure protocol that is based on fundamental quantum mechanics and is not vulnerable to these threats.  The primary figure of merit for QKD systems is ability to generate secret bits. However, to date, methods that have been developed to simulate the secret bit rate generation for EQKD systems have been limited by techniques that do not provide a complete description of the quantum state produced by the source.  In this paper, we provide a complete description and comparison of the secret bit rate for continuous-wave and pulsed laser EQKD systems. In particular, we highlight the relevant Poissonian and thermal photon statistics that affect the EQKD secret bit rate and use practical system parameters and configurations to show regimes where one expects optimal performance for each case.

\end{abstract}
 

\maketitle

\section{Introduction}
Entangled Quantum Key Distribution \cite{Ekert} is a protocol to allow two remote parties to generate a shared secret. The key, or random bit string, is shared between the two parties, Alice and Bob, without a third party, Eve, gaining any information about the key. EQKD consists of three parts and three stages. There is a source of entangled two-level systems, generally pairs of entangled photons produced by down-conversion. Experimentally, this has been preformed for both pulsed and continuous-wave (CW) laser systems \cite{exp, exp-pulsed}. Additionally, there are two detectors each for Alice and Bob. The first phase of the protocol consists of data collection. The source emits a pair and sends one half to Alice and one half to Bob. Alice and Bob each make measurements in one of two orthogonal bases. In the second phase they reconcile the measurements, keeping only the data where they measured in the same basis and they both received a signal. They now share a noisy key between them. The final stage is error correction and privacy amplification. They compare information over an authenticated public classical channel and correct all errors calculating a bit-error rate ($BER$) value. It has been shown the Eve's information is bounded by this $BER$. They then calculate the amount of privacy amplification necessary and perform the privacy amplification to make Eve's information negligible.

The formula \cite{nothing} for the amount of secret key generated is 

\begin{eqnarray}
K =   S + D   - (f_{\rm EC} + f_{\rm PA}) \cdot S \cdot  H_2({\rm BER})   \label{rate}
\end{eqnarray}

\noindent where $S$ is the number of sifted bits, $D$ is the number of dark counts, $f_{EC/PA}$ is an efficiency factor greater than $1$ for error correction or privacy amplification resepctivly, and $H_2$ is the binary Shannon entropy. In the rate formula shown above, the $BER$ is the only variable that is difficult to simulate. \cite{lo} has shown simulations for the $BER$ and key rate in a system with a pulsed-source laser, which obeys thermal statistics. Here, we expand that technique from pulsed systems to allow for the analysis of continuous-wave (CW) laser systems as well as all intermediate ratios of thermal and Poissonian statistics. With the expanded technique in place, we compare CW and pulsed systems for a variety of realistic experimental parameters. The paper is organized as follows: section II describes how to simulate the $BER$ for a CW-laser. Section III shows the results and compares the situations where the source is in Alice's enclave or in the middle. Section IV shows how the procedure in section II can be adjusted for any value from a completely thermal state to a completely Poissionian state. Finally, in section V CW-laser systems are compared to pulsed-laser systems.

\section{Calculating the $BER$ for a CW-laser}

To calculate the $BER$ from a CW-laser system, we can use the same method as in \cite{lo}, except with a different initial state due to the fact that we are now considering a CW source. The initial state for a CW-laser over the time period given by the timing window of Alice's and Bob's detectors \cite{gisin} is 

\begin{equation}
\ket{\Psi} = \frac{1}{C^N} \sum^{\infty}_{M = 0} T^M \ket{\overline{M}(N)}. \label{Gisin-State}
\end{equation}

where $\ket{\overline{M}(N)}$ is the unnormalized state of equal superpositions of all states with $M$ excitation pairs and $N$ modes, $N$ is the number of modes available, approximately $\frac{\delta t}{t_{coh}}$, $\delta t$ is the detector timing window, and $t_{coh}$ is the coherence time of the light field. The constants $C,T$ are defined as $C = \text{cosh}(\xi)$, $T = \text{tanh}(\xi)$ with $\xi$ being proportional to the amplitude of the laser pump field.

Expressing the unnormalized state $\ket{\overline{M}(N)}$ in terms of the normalized state $\ket{M(N)}$, requires counting all the ways that $M$ indistinguishable pairs may be placed into $N$ in principle distinguishable modes. The normalization is then the square root of the number of nonnegative solutions to $x_1 + x_2 + \ldots + x_N = M$,

\begin{eqnarray}
\ket{\overline{M}(N)} &=& \sqrt{{M+N-1}\choose{M}} \ket{M(N)} \nonumber \\
&=& \sum_{\substack{ x_i \geq 0 \\ \sum^{N}_{i=1} x_i = M } } \ket{x_1, x_2, \ldots, x_N}.
 \end{eqnarray}

Since the timing window contains many modes, we count m clicks from all modes the same \cite{squash}. We can neglect which mode is which and define $\ket{\overline{M}_{\pi}(\vec{x})}$ as the unnormalized state corresponding to the symmetric polynomial M$_{\pi}$. Where $\pi$ is some partition of $M$ into $N$ pieces (some may be zero) representing a particular solution to the equation above. Expressing $\pi$ as a multiplicity vector\footnotemark:

\footnotetext{The multiplicity vector is one way of describing a partition. The $i^{\text{th}}$ entry in the vector tells how many times the number $i$ appears in the partition. For example, if $\pi \vdash 5 = \{1,1,3\}$ then the multiplicity vector would be $(2,0,1,0,0)$. Another way to represent partitions used in this paper is the Young Diagram. This is a left-justified array of boxes such that no row has a greater number of boxes than the rows above it.  In the above example $\pi \rightarrow$
 \tiny \yng(3,1,1)
 }

\begin{eqnarray}
\braket{\overline{M}_{\pi}(\vec{x})}{\overline{M}_{\pi}(\vec{x})} &=& \frac{N!}{m_1!  m_2!  \ldots  m_M!  (n - l(\pi))!}  \nonumber \\
&=& M_{\pi}(\vec{\bf{1}})
\end{eqnarray}

\noindent where $l(\pi)$ is the length of the partition. Putting it all together

\begin{eqnarray}
\ket{M(N)} &=& \sqrt{\frac{1}{ {{M+N-1}\choose{M}} }} \sum_{\pi \vdash M} \sqrt{\frac{N!}{m_1! m_2! . . m_M! (N - l(\pi))!}} \nonumber \\
&\times&\frac{1}{\prod_{i=1}^{l(\pi)}\pi_i!} M_{\pi}(\hat{a}_{1}^{\dag}, \hat{a}_{2}^{\dag}, \ldots, \hat{a}_{N}^{\dag} ) \ket{0} \nonumber \\ 
&=& \frac{1}{\sqrt{{M+N-1}\choose{M}}} \sum_{\pi \vdash M} \sqrt{M_{\pi}(\vec{\bf{1}})} \ket{M_{\pi}(\vec{x})}
\end{eqnarray}

The $BER$ of the state $\ket{\Psi}$ is a weighted sum of the $BER$s due to $\ket{M(N)}$ which in turn is a weighted sum of the $BER$s due to $\ket{M_{\pi}(\vec{x})}$.

For a state $\ket{M_{\pi}(\vec{x})}$ with $\pi\vdash M$,  $\pi = \pi_1 \pi_2 \ldots \pi_{l(\pi)}$, we calculate the $BER$.  $\displaystyle \ket{M_{\pi}(\vec{x})} = \frac{1}{\sqrt{C}} \sum_{m=0}^{M} \sqrt{A^{\pi}_{m}} \ket{m, M-m}$ where $\ket{m, M-m}$ is the state with $m$ pairs with horizontal polarization and $M-m$ pairs with vertical polarization. $A^{\pi}_{m}$ is the number of distinct ways you can have $m$ horizontally polarized pairs given a state with a partition $\pi$ of $M$ and $\sqrt{C}$ is the normalization.

If $\pi \rightarrow$ \tiny  \yng(4) \normalsize ... (i.e. a thermal state) then $A^{\pi}_{m}$ is the number of nonnegative integral solutions to the equation

\begin{eqnarray}
\pi_1 = m && x_1 \in \{0,1, .., M\} \nonumber \\
&& x_{i > 1} \in \{0\} \label{thermal}
\end{eqnarray}
\noindent $A^{\pi}_{m} = 1$ and $C = M+1$.

If $\pi \rightarrow $ \tiny \yng(1,1,1,1) \normalsize ... (i.e. a Poissionian state) then $A^{\pi}_{m}$ is the number of nonnegative integral solutions to the equation

\begin{eqnarray}
\pi_1 + \pi_2 + \ldots + \pi_{l(\pi)} = m && x_i \in \{0, 1\} \nonumber \\
&& l(\pi) = M \label{poissionian}
\end{eqnarray}
\noindent $A^{\pi}_{m} = {{M}\choose{m}}$ and $\displaystyle C = \sum_{m=0}^{M} {{M}\choose{m}} = 2^M$.

For a general partition $A^{\pi}_{m}$ is the number of  nonnegative integral solutions to the equation

\begin{eqnarray}
\pi_1 + \pi_2 + \ldots + \pi_{l(\pi)} = m && x_i \in \{0, 1, \ldots , \pi_i\} \nonumber \\
&& l(\pi) = M
\end{eqnarray}
\noindent To find $A^{\pi}_{m}$ let $S$ be the set of nonnegative integral solutions to the unbounded problem. Let $S_i$ be the set of nonnegative integral solutions to the problem where all variables are unbound except $x_i > \pi_i$. Further let $S_A$ be the set of nonnegative integral solutions to the problem where all variables are unbounded except $x_i > \pi_i$,   $\forall i \in A$. Now, we can express $A^{\pi}_{m}$ in terms of $|S|$, $|S_i|$, $|S_A|$.

\begin{eqnarray}
A^{\pi}_{m} &=& |S| - \left|\bigcup_{i=1}^{l(\pi)} S_i\right|   \nonumber \\
&&\text{and by the principle of inclusion and exclusion \cite{combin}} \nonumber \\
A^{\pi}_{m} &=& |S| + \sum_{r=1}^{l(\pi)} (-1)^r \sum_{\substack{ T_r:  |T_r| = r \\ T_r \subseteq \{ 0,1, \ldots, l(\pi)\} } } |S_{T_r}| \nonumber \\
C &=& \sum_{m=0}^{M} A^{\pi}_{m}.
\end{eqnarray}

\noindent Calculating the cardinalities yields

\begin{eqnarray}
|S| &=& {{l(\pi) - 1 + M}\choose{l(\pi) - 1}} \nonumber \\
|S_i| &=& {{l(\pi) - 1 + M - \pi_i - 1}\choose{l(\pi) - 1}} , m \geq \pi_i +1  \nonumber \\
&& 0 \text{   , otherwise} \nonumber \\
|S_A| &=& {{l(\pi) - 1 + M - \sum_{i \in A} (\pi_i + 1)}\choose{l(\pi) - 1}} , m \geq \sum_{i \in A} (\pi_i + 1) \nonumber \\
&& 0 \text{  , otherwise} \nonumber \\
\end{eqnarray}

Now, we have a way of calculating the $BER$ of the state $\ket{M_{\pi}(\vec{x})}$ as a weighted sum of the $BER$ of the states $\ket{m, M-m}$. From \cite{lo} the $BER$ for a state $\ket{m, M-m}$ is

\begin{eqnarray}
e_{mM} &=& e_0 - (\frac{(e_0 - e_d)}{Y_n}\left[ (1 - \eta_A)^{m} - (1-\eta_A)^{M-m} \right] \nonumber \\
&\times& \left[ (1 - \eta_B)^{m} - (1-\eta_B)^{M-m} \right]),
\end{eqnarray}
\noindent where $e_0$ is the error rate of the dark counts ($\frac{1}{2}$), $e_d$ is the detector error rate, and $\eta_{A/B}$ is the detector efficiency of Alice or Bob including the losses in the optical path from the source to the detector.

\section{CW-laser Simulations}

\begin{figure}[tp]
\includegraphics[width=\columnwidth]{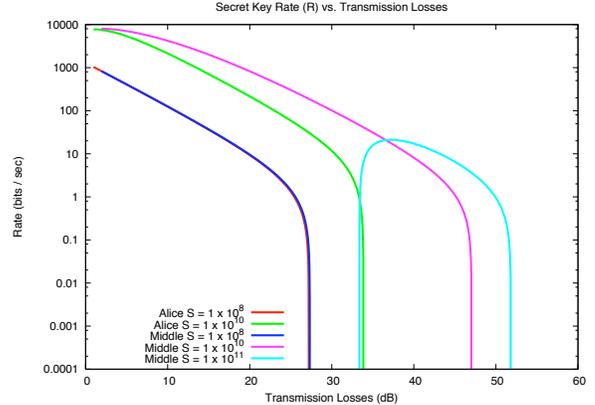}
\caption{A semi-log plot of the secret key generation rate (Eqn. \ref{rate}) vs. the distance for a variety of pair generation rates (S) and for the cases where the source is kept either in Alice's enclave or in the middle. For low strength there is not much difference between the two situations, but at high power the case where the source is in the middle is superior.}
\label{graph_AMdistance}
\end{figure}

\begin{figure}[tp]
\includegraphics[width=\columnwidth]{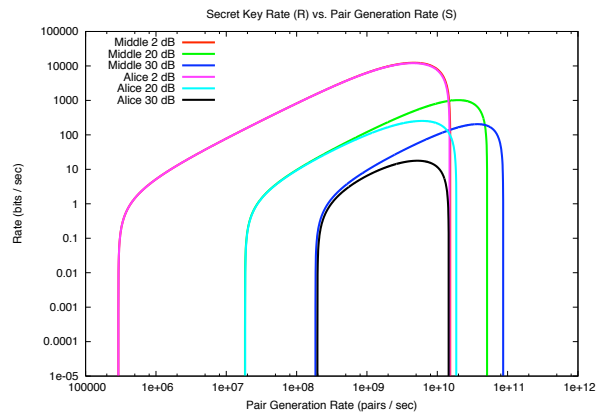}
\caption{A log-log plot of the secret key generation rate (Eqn. \ref{rate}) vs the pair generation rate for a variety of transmission losses for the cases where the source is in Alice's enclave and when it is in the middle. The maximum pair generation rate cutoff is similar for all cases where the source is in Alice's enclave, but increases the losses when the source is placed in the middle.}
\label{graph_AMS}
\end{figure}

With the $BER$ of a state in the form of (Eqn. \ref{Gisin-State}), we can easily simulate what the secret key generation rate will be for any set of system parameters. The results are shown in Figs. \ref{graph_AMdistance} and \ref{graph_AMS}. For the simulations we used a system visibility of $97\%$ and operated for $100$ s and set the optical loss within both Alice's and Bob's enclaves as $7$ dB including detector efficiency. The simulation parameters for the detectors were that they had a dead time of $1\ \mu$s and a dark and background count rate of $1500$ Hz. The timing window was set at $1$ ns. For the source, we used a variable $S$ for the pair generation rate per second. This was modified by a geometric factor of $G = (3\%)^2$, to get the number of pairs sent to Alice and Bob per second. The coherence time for the down-converted light was $100$ femtoseconds.

Fig. \ref{graph_AMdistance} shows the secret key generation rate vs distance between Alice and Bob for a variety of power levels for a system with the source in the middle and for a system with the source inside Alice's enclave. With low power, having the source in the middle doesn't improve the maximum rate significantly. Yet, when $S = 10^{10}$ pairs / s, having the source in the middle improves the distance by $50$ km. Another curious part of the plot is the line for  $S = 10^{11}$ pairs / s and the source in the middle. The key rate is zero for low distances. Counter-intuitively, as the loss in the channel increases the key rate jumps. This is caused by the fact that at lower losses, the probability that a single timing window will have many photons at both Alice's and Bob's side is too high and thus the $BER$ is too high. It is the ability of CW systems to change the effective mean photon number by changing the timing window, combined with locating the source in the middle, that will give the CW system increased flexibility and further distances than a comparable pulsed system. As the losses increase, this problem disappears. When the source is in Alice's enclave, the high power will continue to reach her detector regardless of the distance, so the same phenomenon does not occur. In fact, for $S = 10^{11}$ pairs / s there is no key generated for the system with the source inside Alice's enclave.

The second plot (Fig. \ref{graph_AMS}) shows the secret key generation vs. the pair generation rate. Here, one can see the power cutoff for a particular distance more strikingly. Notice that for the system with the source inside Alice's enclave all the different distances have a similar cutoff, but for the system with the source in the middle, the cutoffs increase with distance.

\section{Comparing CW and Pulsed laser systems}

The approximation that $N = \frac{\Delta t}{t_{\text{coh}}}$ is an integer and thus that each timing window consists of an integral number of temporal modes is only justified for $\frac{\Delta t}{t_{\text{coh}}} >> 1$. It may be useful to use the results of section 2 when this is no longer the case. One can see that for a pulsed laser with $\frac{t_{\text{coh}}}{t_{\text{pulse}}} >> 1$, setting $N = 1$ in section $2$ should reproduce the pulsed laser system. Putting both of the systems in the same framework allows for easy comparisons between them. 

So when $N=1$, the resultant state may only have one form of Young diagram, the horizontal (\ref{thermal}). As $N \rightarrow \infty$, the resultant state has a Young diagram that approaches vertical (\ref{poissionian}). For the mostly likely error, the 4-photon state, the Young diagrams are just \tiny \yng(2) \normalsize and \tiny \yng(1,1) \normalsize .  A Calculation of the $BER$ yields $16\frac{2}{3}\%$  and $25\%$ for the thermal and Poissionian states, respectively. However, that doesn't mean that Poissionian states are worse for QKD. The other factor is with what probability these multi-pair states occur. To compare the mostly Poissionian CW-laser case and the thermal pulsed-laser case, we used the same system parameters as in section 3. With the addition of the fact that for the pulsed system $N=1$ and $t_{\text{pulsed}} = \Delta t = 1$ ns, and $S_{pulsed} = \frac{\mu_{pulsed}}{G \cdot t_{pulsed}}$, so that we can perform a comparison.

\begin{figure}[tp]
\includegraphics[width=\columnwidth]{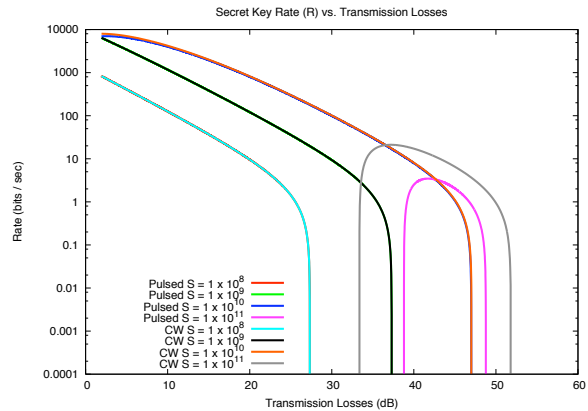}
\caption{A semi-log plot of the secret key generation rate (Eqn. \ref{rate}) vs the distance for a variety of pair generation rates, for both the CW and the pulsed systems, for the case where the source is in the middle. The results for the two systems lie on top over each other for all $S < 1 \times 10^{11}$ pairs / s.}
\label{graph_CWPdistance}
\end{figure}

\begin{figure}[tp]
\includegraphics[width=\columnwidth]{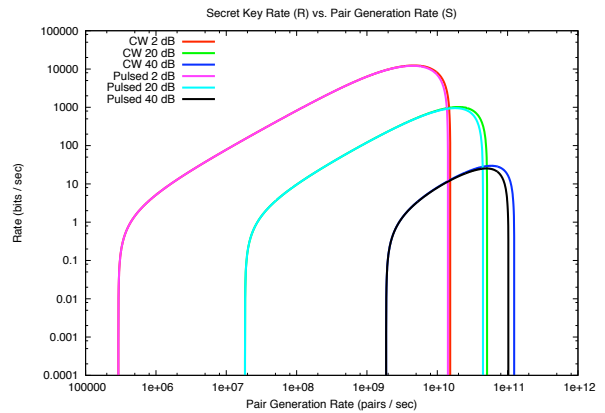}
\caption{A log-log plot of the secret key generation rate (Eqn. \ref{rate}) vs the pair generation rate, for a variety of distances and for both the CW and the pulsed systems when the source is in the middle. For low pair generation rates the two are the same, but when the rates are high, the cutoff is slighty higher for the CW system. This separation increases with the transmission losses.}
\label{graph_CWPS}
\end{figure}

The third plot is again a plot of the secret key rate vs the distance for many pair generation rate $S$. Both systems had a source in the middle. The pulsed and CW systems are mostly identical except in the case of high power. The low loss cutoff is present for both systems, but is worse for the pulsed system. Also with high power the CW system has a positive rate for distances quite a bit further than pulsed systems. The fourth plot shows both that the improvement in cutoff as distance increases for both systems and that the separation between the cutoff for the pulsed and the CW systems increases as distance increases.

\section{Conclusions}

We have presented a method for calculating the expected $BER$ for a CW-laser EQKD system. The method consists of three reductions from the state of the light to easier to handle superpositions of states. First, the light state reduction proceeds from the laser state to a superposition of states with definite excitation number, then to a superposition of partition states. And finally, from a partition state to a superposition of polarization states. The computationally intensive step is calculating all the weights by means of a loop over all possible partitions of a positive integer. We find that for lower pair generation rate (power) the CW and pulsed systems are very close in terms of secret key generation rates and maximum losses. When the pair generation rates are high (much higher than currently available) there is an advantage to using CW laser systems, if all other experimental parameters for the two systems are the same. In addition, including both Poissonian and thermal statistics allows for more practical simulations of EQKD systems.

%\begin{figure}[tp]
%\includegraphics[width=\columnwidth]{paper_N3.pdf}
%\caption{Semi-log plot of the probability that the high strength (key-encoding) decoy state. As the number of counts due to the weaker states increase and the bounds get tighter, a higher fraction of the total signals can be used on the high strength state. $y_0 = 2\times 10^{-6}$, $\epsilon = 10^{-7}$, $V = 98\%$, channel loss $= 30$ dB.}
%\label{graph_N3}
%\end{figure}

%quantum teleportation \cite{bennett93}, quantum

%\bibitem{bennett93} C. H. Bennett \textit{et al} 1993 \textit{Phys. Rev.
%Lett.} \textbf{70 }1895.
%
%\bibitem{devaney} R. L. Devaney, \textit{An Introduction to Chaotic
%Dynamical Systems}, Second Edition, Addison-Wesley, Redwood City, Calif.
%(1989).
%
%\bibitem{norton1} C. J. Villas-Boas, N. G. de Almeida and M. H. Y. Moussa
%1999 \textit{Phys. Rev. A} \textbf{60} 2759.

%\bibliography{DecoyQKD}

\end{document}